\documentclass[%
 lengthcheck,
 amsmath,amssymb,
 aps,
 prl,
 superscriptaddress,
 longbibliography,
 floatfix,
]{revtex4-2}
\usepackage{graphicx}
\usepackage{bm}
\usepackage{xcolor}
\usepackage{xspace}
\usepackage{xfrac}
\usepackage{physics}
\usepackage{microtype}
\usepackage[pdfusetitle,colorlinks,allcolors=blue]{hyperref}

\def\equationautorefname#1#2\null{Eq.#1(#2\null)}

\newcommand{\opEp}{\hat{E}^{+}}

\newcommand{\dw}{\mathrm{d}\omega}
\newcommand{\dt}{\mathrm{d}t}

\newcommand{\FTMCIFIMAC}{%
\affiliation{Departamento de Física Teórica de la Materia Condensada, Universidad Autónoma de Madrid, E-28049 Madrid, Spain}
\affiliation{Condensed Matter Physics Center (IFIMAC), Universidad Autónoma de Madrid, E-28049 Madrid, Spain}
}

\begin{document}

\preprint{APS/123-QED}

\title{Can Intense Quantum Light Beat Classical Uncertainty Relations?}

\author{Felipe R. Willemann}
\email{felipe.reibnitz@uam.es}
\FTMCIFIMAC

\author{Mauro Antezza}
\email{mauro.antezza@umontpellier.fr}
\affiliation{%
 Laboratoire Charles Coulomb (L2C), UMR 5221 CNRS-Université de Montpellier, F-34095 Montpellier, France
}%
\affiliation{%
 Institut Universitaire de France, 1 rue Descartes, Paris Cedex 05, F-75231, France
}%

\author{Johannes Feist}
\email{johannes.feist@uam.es}
\FTMCIFIMAC

\date{December 10, 2025}

\begin{abstract}
Uncertainty relations are fundamental to quantum mechanics, encoding limits on the simultaneous measurement of conjugate observables.
Violations of joint uncertainty bounds can certify entanglement---a resource critical for quantum information protocols and increasingly relevant in strong-field physics.
Here, we investigate the pairwise time-delay and frequency-bandwidth uncertainties for arbitrary multimode quantum states of light, deriving a general lower bound for their joint product.
We find that the nonclassical correction scales inversely with the average photon number, a behavior rooted in the so-called ``monogamy of entanglement''.
These results clarify the intensity scaling of quantum advantages in nonclassical light states and highlight the interplay between entanglement and photon statistics.
\end{abstract}
  
\maketitle

Nearly a century ago, Heisenberg published his fundamental work on the uncertainty principle~\cite{heisenberg1927}.
Originally formulated for the position and momentum of a single particle, it has since been shown to apply to many other systems and pairs of conjugate observables.
Its time-frequency counterpart, the Gabor limit $\Delta t\Delta\omega\geq1/2$~\cite{gabor1946}, is a general property of waves---quantum mechanical or not---which describes how their simultaneous localization in time (duration) and frequency (bandwidth) is fundamentally constrained.
For a two-particle state, one can define the uncertainty in the time delay $\tau = t_1 - t_2$ and in the sum of frequencies $\Omega = \omega_1 + \omega_2$.
The Mancini separability criterion~\cite{mancini2002,howell2004} states that a violation of the joint uncertainty bound $\Delta\tau\Delta\Omega\geq1$ (henceforth denoted the ``classical bound'') is a direct consequence of the presence of quantum correlations between the particles.
As such, the product goes to zero when they are maximally entangled.
While widely used to certify photonic entanglement, generalizing this criterion to fully characterize multiphoton states is nontrivial: even for three photons, a set of more complex inequalities is required~\cite{coelho2009,shalm2013}.

On the other hand, tremendous advancements have been made during the last decades in testing the limits of quantum theory.
In the case of strong-field physics, most phenomena are well described semiclassically, treating light as a classical field~\cite{hentschel2001,corkum2007,krausz2009}.
However, recent work has revisited this assumption, exploring the quantum state of light in intense interactions and revealing nonclassical effects both theoretically~\cite{gonoskov2016,gorlach2020,gombkoto2021,eventzur2023,stammer2023,gonoskov2024,yi2025} and experimentally~\cite{tsatrafyllis2017,lewenstein2021,theidel2024,theidel2025}.
Moreover, the development of bright squeezed vacuum (BSV) sources~\cite{iskhakov2009,spasibko2012,chekhova2015} has enabled applications in strong-field physics, where such states can drive highly nonlinear processes~\cite{gorlach2023,rasputnyi2024,liu2025,mouloudakis2018,mouloudakis2019,lemieux2025,tzur2025}.
These advances push the frontiers of macroscopic quantum mechanics, paving the way towards novel quantum technologies and raising fundamental questions about ultrashort laser pulses~\cite{bhattacharya2023,stammer2023,cruz-rodriguez2024}.
Previous experimental works have successfully proved the resolution advantage of biphotons (entangled photon pairs) in the high-gain regimes of four-wave mixing~\cite{kumar2021} and spontaneous parametric down-conversion (SPDC)~\cite{cutipa2022}.
In the case of spatio-temporal correlations created in high-gain SPDC, the biphotonic wavefunction is modeled to describe the so-called ``\emph{X}-entanglement'' present in macroscopic twin beams~\cite{brambilla2010,brambilla2012,sharapova2020} and measurements of the coherent contribution of entangled photon pairs were possible through sum-frequency generation.

In this work, we investigate a generalization of the time-frequency uncertainty relation to answer a fundamental question: can intense quantum light have a general time-frequency resolution advantage over classical fields?
It is well-known that no such advantage exists for the overall duration $\Delta t$ and frequency bandwidth $\Delta\omega$ of a pulse, as these are single-particle observables for which the Gabor limit is valid~\cite{wigner1988,busch2002}.
In contrast, up to now no general bound for the joint uncertainty product $\Delta\tau\Delta\Omega$ for arbitrary photon states has been reported.
For two-photon states, it is clear that the classical bound $\Delta\tau\Delta\Omega\geq 1$ can be violated as a consequence of quantum correlations, but it is not known how this extends to states with a higher number of photons---for which the observables $\Delta\tau$ and $\Delta\Omega$ include the contribution of all photon pair permutations in the state. 
We here show that the Mancini criterion still applies for such a generalization of the uncertainty product and derive a general bound for nonclassical states, given in simplified form by
\begin{equation}
    \Delta\tau\Delta\Omega \geq \sqrt{1-\frac{2}{\expval*{n}}},
    \label{eq:global_bound}
\end{equation}
where $\expval*{n}$ is the average number of photons in the field.
This expression is a simplification of a more general result given below, and is valid when $\expval*{n}\geq2$ and the two-photon component of the quantum state is negligible compared to the average number of photon pairs, as holds for intense pulses.
This new bound clearly shows that although a quantum advantage in the two-photon delay-energy uncertainty is possible, it becomes increasingly small as the intensity increases due to the inverse proportionality to the average number of photons.

Previous works have demonstrated improved time-frequency resolution in intense pulses~\cite{dayan2004,schlawin2018,cutipa2022}, but these relied on measurements sensitive only to the ``coherent'' biphoton contribution in BSV from SPDC---effectively selecting photons produced as entangled pairs.
In generic measurements where all photon pairs contribute, the global joint uncertainty $\Delta\tau\Delta\Omega$ is the relevant quantity, determining, e.g., the total-energy and relative-temporal resolution in two-photon double ionization~\cite{feist2009,pazourek2015,carpeggiani2019,chattopadhyay2023,arteaga2024,hell2025}.

First, let us provide some intuition of the classical bound $\Delta\tau\Delta\Omega\geq 1$.
Since photons in classical fields are statistically independent, the uncertainty in the time delay $\Delta\tau$ between them is directly determined by the pulse duration (i.e., uncertainty in arrival time) $\Delta t$, and given by $\Delta\tau^2 = 2\Delta t^2$.
Thus the only way to change the average delay between photons in classical electromagnetic fields is by changing the duration of the pulse itself.
Similarly, the uncertainty in the sum of frequencies $\Delta\Omega$ is determined by the combined uncertainties of each photon, and given by $\Delta\Omega^2 = 2\Delta\omega^2$.
The joint uncertainty product $\Delta\tau\Delta\Omega = 2\Delta t\Delta\omega$ is thus bounded by the Gabor limit.
This means that, for classical fields, the only way to reduce the uncertainty in the time delay between photons is by increasing the bandwidth of the pulse, and vice versa.
This implies that in any process that depends on both the delay and the total energy of all pairs of photons, such as two-photon double ionization, there is a trade-off between temporal and spectral resolution.
In such processes, controlling the time delay can be used to control and investigate electron correlations, but the loss of energy resolution prevents a complete characterization of the process for small delays~\cite{feist2009,pazourek2015,carpeggiani2019,chattopadhyay2023,arteaga2024,hell2025}.

In order to extend this analysis to nonclassical fields, we first obtain an $n$-dependent lower bound for pure quantum states of light within a subspace of fixed photon number $n$.
We then generalize this bound to arbitrary (pure or mixed) states containing contributions with different numbers of photons, with a result that for large average photon number $\expval{n}$ simplifies to \autoref{eq:global_bound}.
\begin{figure}
    \centering
    \includegraphics[width=1.0\linewidth]{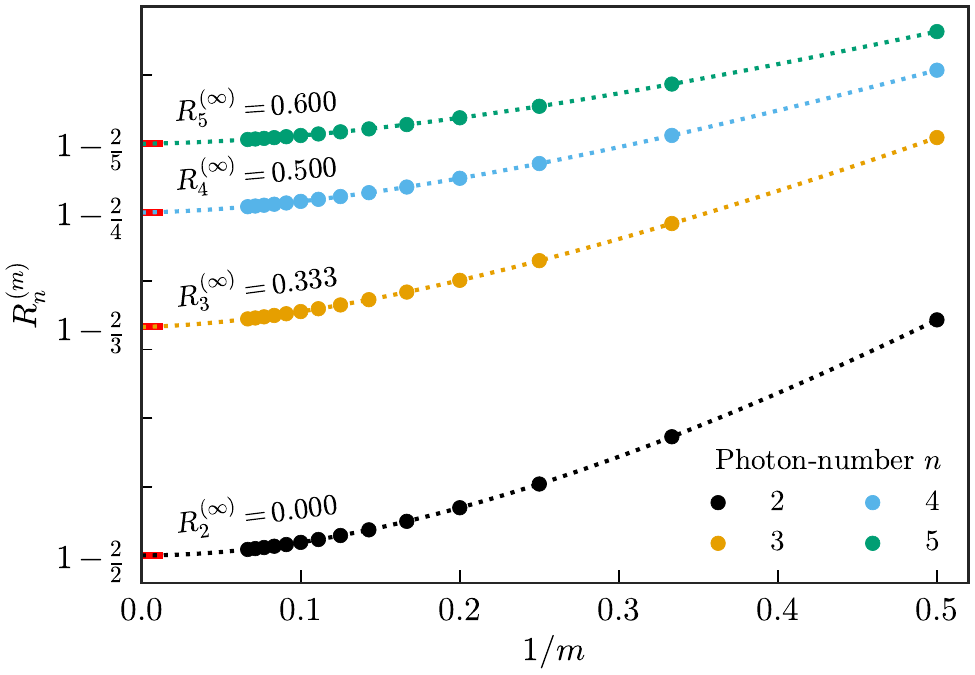}
    \caption{Minimum uncertainty values of the joint time-bandwidth uncertainty product calculated with the uncertainty Hamiltonian method~\cite{opatrny1995} for states with $n\in[2,5]$ photons and Hermite-Gauss mode expansions with $m\in [2,15]$ modes. Dashed lines correspond to fits up to 6th order in $m^{-1}$, with $R_n^{(\infty)}$ the estimated value for infinite basis set size.}
    \label{fig:conver_joint_gs}
\end{figure}

We start by considering the joint uncertainty product of pure states within a fixed $n$-photon subspace, $\Delta\tau_n\Delta\Omega_n$, measured at the same point in space.
We do not impose any restriction on the kind of correlations present in the state or the type of state, but consider the most general case of an arbitrary multimode state of $n$ photons.
The variances of the time delay and sum of frequencies can then be expressed as (see Appendix A for details)
\begin{align}
    \label{eq:rel_frac_tau}
    \frac{\Delta{\tau^2_n}}{2\Delta t^2_n} &= 1 - \frac{\iint\dt\dt' \expval{\hat{a}^\dagger(t)\hat{a}^\dagger(t')\hat{a}(t')\hat{a}(t)} tt'}{(n-1) \int\dt \expval{\hat{a}^\dagger(t)\hat{a}(t)}t^2}, \\
    \frac{\Delta{\Omega^2_n}}{2\Delta\omega^2_n} &= 1 + \frac{\iint\dt\dt' \expval{\partial_{t}\hat{a}^\dagger(t)\hat{a}^\dagger(t')\partial_{t'}\hat{a}(t')\hat{a}(t)}}{(n-1) \int\dt \expval{\partial_t\hat{a}^\dagger(t)\partial_t\hat{a}(t)}},
    \label{eq:rel_frac_omega}
\end{align}
where $\hat{a}(t)$ and $\hat{a}^\dagger(t)$ are continuous-mode annihilation and creation operators of the EM field and the expressions are given in the frame of reference of the pulse where $\expval{t}=\expval{\omega}=0$.
Notice that, if the state is separable, the second-order correlation functions in the second terms of \autoref{eq:rel_frac_tau} and \autoref{eq:rel_frac_omega} factorize over $t$ and $t'$.
Then, the double integrals become proportional to $\expval{t}^2$ and $\expval{\omega}^2$, respectively, which are identically zero by construction, and the classical limit is recovered.
This implies that any possible quantum advantage arises purely from entanglement, i.e., nonseparability of the state. Therefore, the Mancini criterion still certifies entanglement, though for $n>2$ it is not sufficient to fully characterize the specific type of entanglement.

Although this analysis is instructive, it does not provide direct information on the lower bound of the joint uncertainty product.
Since deriving a general bound analytically is not trivial, we obtain this bound numerically by minimizing an appropriately chosen ``uncertainty Hamiltonian''~\cite{opatrny1995} (see Appendix B for details).
To do so, we expand the field operators in temporal Hermite-Gauss modes and calculate the minimum uncertainty product $R_n^{(m)} = \Delta\tau^2_n\Delta\Omega^2_n$ for an $n$-photon state and a basis set size of $m$ modes, with results shown in \autoref{fig:conver_joint_gs}.
We then obtain an estimate for the converged infinite-basis-size limit $m\to\infty$ by fitting the numerical results to an asymptotic expansion up to sixth order in $m^{-1}$, i.e., $R_n^{(m)} \approx R_n^{(\infty)} + \sum_{i=1}^6 a_i {m}^{-i}$.
Here, $a_i$ are constants and $R_n^{(\infty)} \approx R_n$ is the estimated minimum uncertainty product for the $n$-photon state when using an infinite (exact) expansion.
It can be seen that these estimates follow, up to a very good numerical approximation, the law
\begin{equation}
    \Delta\tau^2_n\Delta\Omega^2_n \geq 1-\frac{2}{n}.
    \label{eq:limit_nsub}
\end{equation}

While we are not aware of an analytical proof for arbitrary states, we note that this limit can be obtained analytically for states with joint temporal amplitudes $\phi_n(\boldsymbol{t})$ of the exchange-symmetric multivariate Gaussian form
\begin{equation}
    |\phi_n (\boldsymbol{t})|^2 = A_n e^{-\frac{\gamma^2}{n}\left(\sum\limits_{i=1}^n t_i\right)^2 - \frac{\delta^2}{n} \sum\limits_{i<j} (t_i-t_j)^2},
    \label{eq:multivar_gauss}
\end{equation}
where $A_n$ is a normalization constant and $\gamma$ and $\delta$ are positive real parameters whose ratio defines the kind of correlation present in the state.
For $n=2$, this distribution has been used to model colinear type-II SPDC, in which $\gamma$ represents the pump bandwidth and $\delta$ is the inverse of the so-called ``entanglement time'' between photons~\cite{grice1997,lavolpe2021}.
The uncertainty product for such states has the closed form
$\Delta\tau^2_n\Delta\Omega^2_n = 1 - \frac{2}{n} (1 - \frac{\gamma^2}{\delta^2})$.
For $\gamma=\delta$, the state is separable and the classical bound is obtained as expected.
When $\gamma < \delta$ (corresponding to positive temporal correlation), the uncertainty product goes below the classical bound, which, again, is a clear indicator of nonclassicality.
Moreover, as $\gamma / \delta \to 0$, the uncertainty product reaches the minimum possible for this kind of state, which coincides exactly with \autoref{eq:limit_nsub}.
We also note that this statement does not depend on the shape of the overall temporal envelope (given by the first exponential term in \autoref{eq:multivar_gauss}), as any function converging to a constant as a function of $(\sum_i t_i)^2$ will yield the same limiting value.

We next provide a theoretical argument that suggests that densities of the form of \autoref{eq:multivar_gauss} indeed provide the minimum-uncertainty states.
Applying the Cauchy-Schwarz inequality to the uncertainty product in the photon-number subspace expressed in terms of the joint temporal amplitude gives 
\begin{equation}
    \Delta\tau^2_n\Delta\Omega^2_n \geq \bigg|\!\int\!\dd[n]{\boldsymbol{t}} (t_i-t_j)\phi^*_n(\boldsymbol{t})(\partial_{t_i}+\partial_{t_j})\phi_n(\boldsymbol{t})\bigg|^2
\end{equation}
where $i$ and $j$ are any two distinct indices in $\{1,\ldots,n\}$.
Due to bosonic symmetry, $\phi_n(\boldsymbol{t})$ is symmetric under exchange of any two indices, such that the integrand is antisymmetric under exchange of $i$ and $j$, and the integral evaluates to zero.
This would imply that the uncertainty product can be made arbitrarily small.
However, for equality in the Cauchy-Schwarz inequality to be achieved, the functions must be linearly dependent, which here implies $(\partial_{t_i}+\partial_{t_j})\phi_n(\boldsymbol{t})\propto(t_i -t_j)\phi_n(\boldsymbol{t})$.
Due to the bosonic exchange symmetry, derivation with respect to one index must give a result that is symmetric in all other indices, such that this can only be fulfilled for $n=2$, specifically when $\partial_{t_1}\phi_2(\boldsymbol{t})\propto(t_2-t_1)\phi_2(\boldsymbol{t})$, such that the sum of derivatives is zero.
Extending this condition to $n\geq 3$ while enforcing bosonic symmetry implies
\begin{equation}
    \partial_{t_i}\phi_n(\mathbf{t}) \propto \sum_{k\neq i}^n(t_k-t_i)\phi_n(\mathbf{t}),
    \label{eq:gauss_condition}
\end{equation}
and it is easy to see that the general solution are Gaussian wavefunctions whose squares are the densities of \autoref{eq:multivar_gauss} for $\gamma=0$, giving the bound of \autoref{eq:limit_nsub}.

This argument, supported by our numerical results, provides strong evidence that the lower bound for the uncertainty product for pure states with a given photon number is indeed given by \autoref{eq:limit_nsub}.
This lower bound can also be related to the well-established concept of monogamy of entanglement~\cite{coffman2000,adesso2006,adesso2006a}, since as discussed before, deviation from the classical value of the joint uncertainty is solely due to time-energy entanglement.
This concept establishes that if a degree of freedom---in the present case, the arrival time of a photon, or its detected frequency---is maximally entangled with another one, then it cannot be simultaneously maximally entangled also with a third one.
For a two-photon state, there is only bipartite entanglement, which is not limited, thus the uncertainty relation can go to zero.
For states with a higher number of photons to reproduce this limit, all pairs of photons should exhibit maximum mutual entanglement, since the observables are sensitive to all pairs of photons.
However, the monogamy principle limits how much quantum correlation can be shared between all particles, and in turn how correlated each pair can be.
Note that although our derivation of \autoref{eq:gauss_condition} relies on the fact that all the photons in the field are indistinguishable, the conceptual arguments given above are also applicable to indistinguishable particles.

We now generalize our result to arbitrary (mixed or pure) states containing contributions with different photon numbers.
The observables $\Delta\tau$ and $\Delta\Omega$ are diagonal in the photon number $n$, such that the contribution of each photon-number subspace to each observable is independent, and coherences between different photon-number subspaces do not matter.
For an arbitrary quantum state described by density matrix $\hat{\rho}$, we then get 
\begin{equation}
    E(\hat{\rho}) = \frac{\expval{n(n-1) E(\hat{\rho}_n)}}{\expval{n(n-1)}},
\end{equation}
where $E$ is either $\Delta\tau^2$ or $\Delta\Omega^2$, and we used the brackets to denote the statistical average over the photon-number distribution $\expval{x} = \sum_n p_n x_n$.
Here, $p_n$ is the probability of measuring $n$ photons, with $\sum_n p_n = 1$ and mean photon number $\expval{n}=\sum_{n}p_n n$, and $\hat{\rho}_n$ is the normalized density matrix in the $n$-photon subspace.
The factors $n(n-1)$ give the number of pairs of photons in each subspace, and appear because we are considering two-photon observables.
We can further expand $\hat{\rho}_n = \sum_i p_{ni} \ket{\phi_{ni}}\bra{\phi_{ni}}$, where $\ket{\phi_{ni}}$ are (normalized) pure states in the $n$-photon subspace and $p_{ni}$ are their corresponding probabilities, such that $\sum_i p_{ni} = 1$.
In terms of these states, we have $E(\hat{\rho}_n) = \sum_i p_{ni} E(\ket{\phi_{ni}})$ (since $n$ is constant, the factor for number of pairs cancels away).
For each pure state $\ket{\phi_{ni}}$, the uncertainty product is bounded by \autoref{eq:limit_nsub}, i.e., $\Delta\tau_{ni}^2\Delta\Omega_{ni}^2 \geq 1 - \frac{2}{n}$.
Using the Cauchy-Schwarz inequality, it is then straightforward to show that $\Delta\tau_n^2 \Delta\Omega_n^2 \geq 1 - \frac{2}{n}$.

Now considering the full state, a lower bound for the joint uncertainty product can be obtained as
\begin{align}
    \Delta\tau\Delta\Omega&=\sqrt{\frac{\expval*{n(n-1)\Delta\tau^2_n}}{\expval*{n(n-1)}}\frac{\expval*{n(n-1)\Delta\Omega^2_n}}{\expval*{n(n-1)}}}\nonumber\\
    &\geq \frac{\expval*{n(n-1)\Delta\tau_n\Delta\Omega_n}}{\expval*{n(n-1)}}\geq\frac{\expval{n(n-1)\sqrt{1-\frac{2}{n}}}}{\expval*{n(n-1)}}\nonumber\\
    &\geq \sqrt{1-2\frac{1-p_0-p_1-p_2}{\expval*{n}-p_1-2p_2}}\left(1-\frac{2p_2}{\expval*{n(n-1)}}\right)
    \label{eq:general_uncert_limit}
\end{align}
where $p_0$, $p_1$ and $p_2$ are the probabilities of measuring the state with zero, one and two photons, respectively.
Here, we used first the Cauchy-Schwarz inequality and then Jensen's inequality in its probabilistic form.
If, for simplicity and given that we are interested in understanding intensity scaling, the two-photon state is assumed to have negligible probability in comparison to the average number of pairs $\expval*{n(n-1)}/2$, after some further simplifications (see Appendix C) for $\expval*{n}\geq2$ we are left with \autoref{eq:global_bound}.
Then, it is clear from \autoref{eq:general_uncert_limit} that the joint time-frequency resolution advantage provided by entanglement is, in general, indeed inversely proportional to the mean photon-number to leading order.
For pulses of high intensity with $\expval*{n}\gg 1$, no matter the form that such nonclassical correlations take, the joint uncertainty is thus governed by the classical bound $\Delta\tau\Delta\Omega\geq1$.
Also note from the second line of \autoref{eq:general_uncert_limit} that any state composed of separable photon-number states immediately reduces to the classical bound, regardless of intensity.
\begin{figure}
    \centering
    \includegraphics[width=1.0\linewidth]{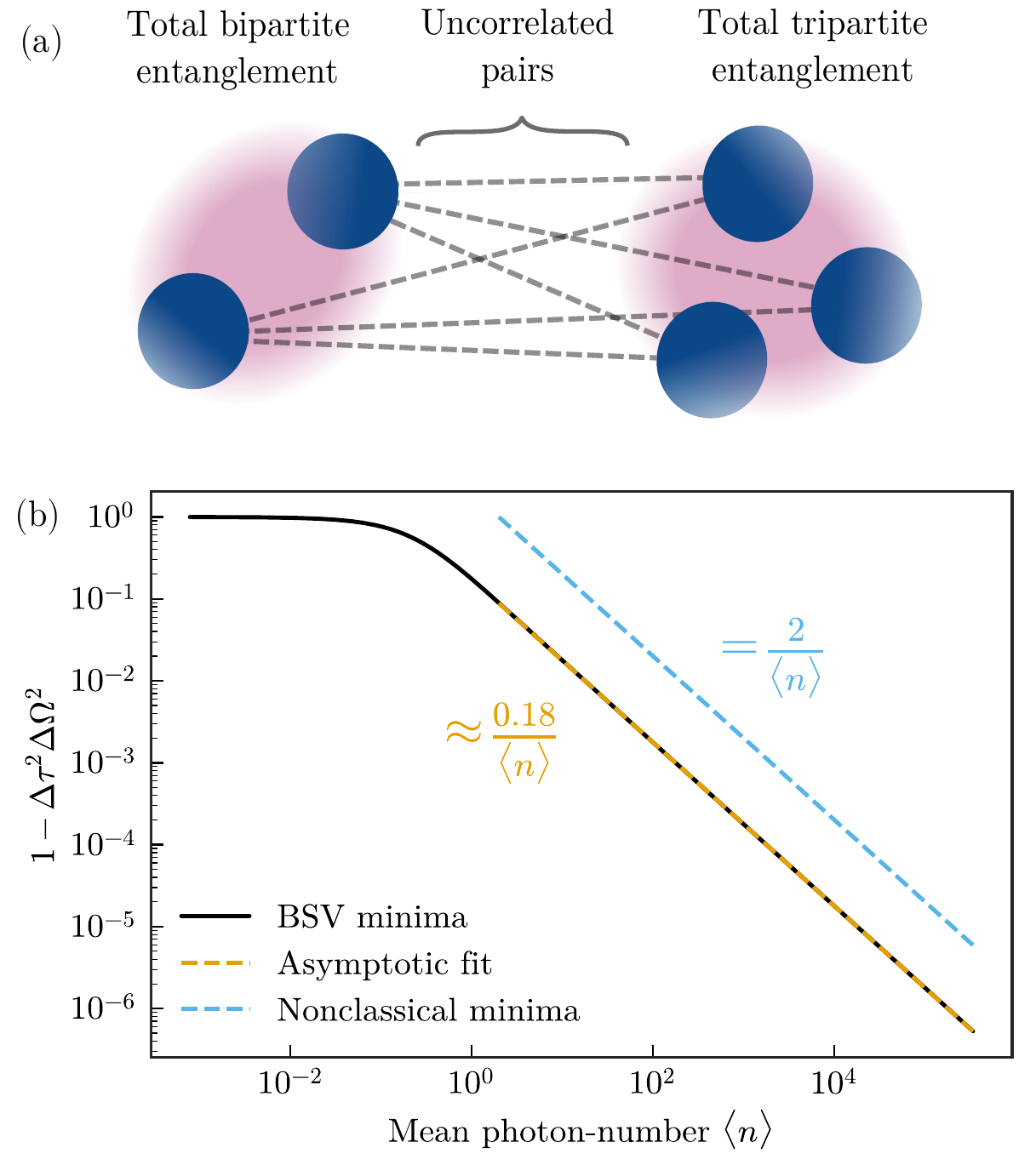}
    \caption{Scaling analysis. (a) Representation of a superposition of two- and three-photon states highlighting the fact that there are overall more uncorrelated than correlated pairs even when each state is individually totally entangled. (b) Negative of the nonclassical correction to the joint time-frequency uncertainty product for the minima of the BSV, with a linear asymptotic fit $\log{(1-\Delta\tau^2\Delta\Omega^2)}=\log{c}-k\log{\langle n\rangle}$ ($k\approx1,c\approx0.18$) and the general nonclassical bound of \autoref{eq:global_bound} for comparison.}
    \label{fig:entangled_pairs}
\end{figure}

The physical interpretation of the loss of advantage in the general cases of superpositions and mixed states is similar to the one usually given for the recovery of classical two-photon absorption rates for BSV~\cite{mouloudakis2018,gea-banacloche1989,dayan2005,lee2006,schlawin2024,raymer2021,gatti2025,dickinson2025}.
As the intensity increases, the observables are dominated by pairs of uncorrelated photons (incoherent contribution) rather than correlated ones (coherent contribution).
This effect is illustrated in \autoref{fig:entangled_pairs}(a) for a superposition of two- and three-photon states, in which there are a total of $4$ correlated pairs compared to $6$ uncorrelated ones.
Moreover, for a superposition of photon-number states ranging from $2$ to $n$ totally entangled photons, it is easy to calculate that the total amount of correlated pairs grows as $n^3$, while that of uncorrelated pairs as $n^4$.
As such, the ratio of correlated to uncorrelated photon pairs is of the order $n^{-1}$, the same scaling we derived for the nonclassical correction to the joint time-bandwidth uncertainty relation.
As a concrete example of an experimentally available state, we study BSV, which is composed of even bi-separable entangled photon states.
Using a colinear version of the SPDC model described in~\cite{schlawin2024}, we calculate the minima of the joint uncertainty product $\Delta\tau^2\Delta\Omega^2$ over the space of squeezing parameters as a function of the mean photon-number.
As shown in \autoref{fig:entangled_pairs}(b), we find that BSV indeed violates the classical bound, with $\Delta\tau^2 \Delta\Omega^2 \approx 1 - \frac{0.18}{\expval{n}}$ for $\expval{n}\gg1$. The deviation thus scales with the minimal order ($\expval{n}^{-1}$) predicted by \autoref{eq:global_bound}, however, with a much smaller prefactor of $0.18$ instead of $2$, such that the time-frequency resolution advantage provided by BSV does not saturate the nonclassical bound.

\emph{Conclusion}---In summary, we showed that while classical and coherent fields are fundamentally limited in the simultaneous resolution of average time delay and bandwidth, nonclassical features in the field can break this limit beyond states composed of biphotons.
We demonstrated a new general lower bound for the joint time-frequency uncertainty product.
The correction to the classical bound was shown to be due to the temporal/spectral nonseparability of light states and, moreover, is shown to be negative, i.e., improve the resolution beyond the classical limit when photon arrival times are positively correlated, as expected.
For states with a fixed photon number in particular, we provided numerical evidence for the minimum bound and theoretical arguments for the $n$-photon density that defines the minimum-uncertainty states saturating the bound.
In these states, the entanglement is equally distributed between all photons, which is a direct consequence of their bosonic symmetry.
Finally, we generalized our results to arbitrary pure and mixed states with contributions from different photon numbers, showing that the possible quantum advantage in the joint uncertainty product decreases inversely with the average photon number.
Our findings provide insights on the possible time-frequency resolution advantage of quantum light in multi-photon processes.
This advantage is found to quickly disappear for intense light with a large mean photon number, a fact that can be understood intuitively through the monogamy property of entanglement.
Since the observed bound depends on the average photon number, this implies that for a given desired average field intensity (which is what effectively determines the rate of $n$-photon processes), the possible quantum advantage can be increased by using modes with smaller effective mode volumes, such as provided by nanophotonic structures~\cite{tame2013,chikkaraddy2016,gonzalez-tudela2024}.

\begin{acknowledgments}
\emph{Acknowledgments}---We thank Frieder Lindel for fruitful discussions.
This work was funded by the Spanish Ministry of Science, Innovation and Universities--Agencia Estatal de Investigación through grants CNS2023-145254, PID2021-125894NB-I00 and PID2024-161142NB-I00, and the ``María de Maeztu'' Programme for Centers of Excellence in R\&D through grant CEX2023-001316-M.
We also acknowledge financial support from the European Union's Horizon Europe research and innovation programme under Grant Agreement No.
101070700 (MIRAQLS).
F.R.W. is grateful to the Theory of Light-Matter and Quantum Phenomena group at the University of Montpellier for their hospitality during his stay in Montpellier, where part of this work was carried out.
\end{acknowledgments}

\nocite{*}
\bibliography{refs}

\appendix
\section{End matter}
\emph{Appendix A: Field observables}---We treat light in free space, measured at the origin and only consider a single polarization direction without loss of generality.
Therefore, we can switch to the basis of ``emitter-centered'' modes~\cite{feist2020}, where only a single mode at each frequency contributes to the field at that point and frequency.
The positive part of the continuous-mode electric field operator in the Heisenberg picture is $\opEp_z(\mathbf{0}, t) = \opEp(t) = \int_0^\infty\dw \mathcal{A}(\omega) \hat{a}(\omega) e^{-i\omega t}$, where $\mathcal{A}(\omega) = \sqrt{\hbar\omega^3/6\pi^2\epsilon_0 c^3}$.
We employ the slowly varying envelope approximation, in which we assume that the bandwidth of the field is much smaller than its central frequency $\omega_0$ and that its spectrum vanishes at the origin.
Then, the mode amplitude can be taken out as a constant $\mathcal{A}(\omega) \approx \mathcal{A}(\omega_0)$ and the integrals can be formally extended to negative frequencies.
The field operator becomes
\begin{equation}
    \opEp(t) = \sqrt{2\pi}\mathcal{A}(\omega_0) \hat{a}(t),
    \label{eq:cont_field}
\end{equation}
where $\hat{a}(t) = \frac{1}{\sqrt{2\pi}}\int\dw \hat{a}(\omega) e^{-i\omega t}$ is the Fourier transform of the continuous-mode annihilation operator and the photon-number operator is $\hat{n}=\int\dt\hat{a}^\dagger(t)\hat{a}(t)$.

The observables of interest are the one-photon time and frequency variances and two-photon delay and bandwidth between photon pairs.
In the continuous-mode formalism, they can be calculated from the normalized temporal intensity and second order correlation function, respectively.
In the time-frequency frame of the pulse where $\expval{t} = \expval{\omega} = 0$ (without loss of generality), we have
\begin{gather}
   \Delta t^2 = \frac{1}{\expval{\hat{n}}} \int\dt \expval{\hat{a}^\dagger(t)\hat{a}(t)} t^2\label{eq:var_t}\\
   \Delta\omega^2 = -\frac{1}{\expval{\hat{n}}} \int\dt \expval{\hat{a}^\dagger(t)\partial_t^2\hat{a}(t)}\label{eq:var_w}\\
   \Delta{\tau}^2 = \frac{\int\dt\dt' \expval{\hat{a}^\dagger(t)\hat{a}^\dagger(t')\hat{a}(t')\hat{a}(t)} (t-t')^2}{\expval{\hat{n}(\hat{n}-1)}}\label{eq:squared_delay}\\
   \Delta{\Omega}^2 = -\frac{\int\dt\dt' \expval{\hat{a}^\dagger(t)\hat{a}^\dagger(t')(\partial_{t}+\partial_{t'})^2 \hat{a}(t')\hat{a}(t)}}{\expval{\hat{n}(\hat{n}-1)}},\label{eq:squared_freq}
\end{gather}
where the spectral uncertainties are obtained from the Fourier transforms using the derivative property of the Dirac delta function.

\emph{Photon-number states}---Now, we consider pure states of fixed number of field excitations, i.e., photons, which are, in general, nonseparable.
Their general form can be written as
\begin{gather}
    \label{eq:number_def}
    \ket{\phi_n} = \frac{1}{\sqrt{n!}} \int \dd[n]{\boldsymbol{t}}\phi_n(\boldsymbol{t}) \prod_{i=1}^n \hat{a}^\dagger(t_i) \ket{0},
\end{gather}
where $\phi_n(\boldsymbol{t})$ is the \emph{n}-photon joint temporal amplitude function.
It is assumed to be symmetric under the exchange of any two time variables without loss of generality, and is normalized as $\int \dd[n]{\boldsymbol{t}} |\phi_n(\boldsymbol{t})|^2 = 1$.
The integrals are performed over the vector $\boldsymbol{t} = (t_1,t_2,\cdots,t_n)\in\mathbb{R}^n$, which comprises all photon arrival time variables in the given state.
Moreover, the states of \autoref{eq:number_def} are eigenstates of the number operator with eigenvalue equal to the number of photons $n$, such that the normalization prefactors reduce to constants.
Then, the quadratic terms of \autoref{eq:squared_delay} reduce to $(n-1) \int\dt \expval{\hat{a}^\dagger(t)\hat{a}(t)} t^2$, where we used the commutation relation of the time-domain annihilation operator $[\hat{a}(t),\hat{a}^\dagger(t')]=\delta(t-t')$.
By comparing the result to \autoref{eq:var_t} (and similarly for the frequency observables) we arrive at \autoref{eq:rel_frac_tau} and \autoref{eq:rel_frac_omega}, in which the nonclassical corrections are given by the crossterms.
Finally, by applying \autoref{eq:number_def} on \autoref{eq:squared_delay} and \autoref{eq:squared_freq} and using bosonic symmetry, we get
\begin{gather}
    \label{eq:taun}
    \Delta\tau^2_n = \int \mathrm{d}^{n}\boldsymbol{t} |(t_i-t_j)\phi_n(\boldsymbol{t})|^2\\
    \label{eq:omn}
    \Delta\Omega^2_n = \int \mathrm{d}^{n}\boldsymbol{t} |(\partial_{t_i}+\partial_{t_j})\phi_n(\boldsymbol{t})|^2,
\end{gather}
where the indexes $i,j\in\{1,\ldots,n\}$ and $i\neq j$ (due to bosonic symmetry, the result does not depend on the choice of $i$ or $j$).
Note that, alternatively, one can use the inverse Fourier transform to obtain all the above expressions in frequency space.

\emph{Appendix B: Numerical details}---In order to calculate the minimum joint uncertainty product for the photon-number states, we define the uncertainty Hamiltonian operator as
\begin{equation}
    \hat{H}_{\mathrm{unc}}(\xi)=\xi\hat{\tau}^2+(1-\xi)\hat{\Omega}^2,
    \label{eq:unc_ham}
\end{equation}
where $\xi\in[0,1]$, $\hat{\tau}^2 = \int\dt\dt' \hat{a}^\dagger(t)\hat{a}^\dagger(t')\hat{a}(t')\hat{a}(t) (t-t')^2$ and $\hat{\Omega}^2 = -\int\dt\dt' \hat{a}^\dagger(t) \hat{a}^\dagger(t') (\partial_{t}+\partial_{t'})^2 [\hat{a}(t') \hat{a}(t)]$.
The ground state of the Hamiltonian is the state that minimizes the product $\langle{\hat{\tau}^2}\rangle\langle{\hat{\Omega}^2}\rangle$, which is proportional to the joint uncertainty product (see Appendix A), for a given $\xi$.
The global uncertainty minima corresponds to the lowest ground state eigenvalue in the domain of $\xi$~\cite{opatrny1995}.

We then rewrite $\hat{\tau}^2$ and $\hat{\Omega}^2$ using a basis of Hermite-Gaussian functions instead of the continuous temporal mode basis~\cite{rohde2007}.
The new mode operators are defined as $a_n = \int_{-\infty}^{\infty} \psi_n(t) \hat{a}(t)$, where $\psi_n(t)$ are the real, orthonormal ($\int \psi_n(t) \psi_m(t) \dt = \delta_{nm}$) and complete ($\sum_n \psi_n(t) \psi_n(t') = \delta(t-t')$) Hermite-Gauss functions.
It can be checked that with the definition above, and using $[\hat{a}(t),\hat{a}^\dagger(t')] = \delta(t-t')$, we get $[a_n,a_m^\dagger] = \delta_{nm}$, and it also follows that the inverse transformation is $\hat{a}(t) = \sum_n \psi_n(t) a_n$.
So we have a unitary transformation between a continuous variable and a discrete index, and $\psi_n(t)$ can be seen as a ``unitary matrix'' with ``indices'' $n$ and $t$.

The implementation of the operators in the discretized basis was done in QuTiP 5~\cite{lambert2024} using the excitation-number restricted representation constrained to a fixed total amount of excitations to enforce that the calculation is being done in a given photon-number subspace.
Finally, to find the global minima for a given total number of modes in the expansion, we employ a 1-D nonlinear optimization routine in the interval of the $\xi$ parameter.

\emph{Appendix C: details on the derivation of \autoref{eq:general_uncert_limit}}---Staring from the full product in the first line of \autoref{eq:general_uncert_limit}, we apply the Cauchy-Schwarz inequality of the form $E[XY]^2\leq E[X^2]E[Y^2]$, where $E$ is the expectation value of any two random variables $X$ and $Y$. In our case, the random variables are functions of the number of photons ($\sqrt{n(n-1)}\Delta\tau_n$ and $\sqrt{n(n-1)}\Delta\Omega_n$) for a given photon-number distribution defined by the set of probabilities $p_n$. The second line of \autoref{eq:general_uncert_limit} is then obtained by applying the nonclassical bound $\Delta\tau_n\Delta\Omega_n\geq\sqrt{1-\frac{2}{n}}$ on the bound given by the Cauchy-Schwarz inequality. 

Next, we use Jensen's inequality, which states that for any convex function $f$ of a random variable $X$ one has $E[f(X)]\geq f(E[X])$. However, since in our case the real function $n(n-1)\sqrt{1-\frac{2}{n}}$ is not defined for $n<2$, we use a slightly different version of the inequality, valid for any finite distribution with weights $a_n$. We define the real function $f(n)=(n-1)\sqrt{1-\frac{2}{n}}$, which is strictly convex only for $n\geq3$, such that the summations start from $n=3$. Then, we take $a_n=p_nn$ as the weights and apply the inequality as
\begin{equation}
    \frac{\sum_{n=3}p_nnf(n)}{\sum_{n=3}p_nn} \geq f\left(\frac{\sum_{n=3}p_nn^2}{\sum_{n=3} p_nn}\right).
\end{equation}

Therefore, the bound on the uncertainty product becomes
\begin{align}
    \Delta\tau\Delta\Omega&\geq\frac{\sum_{n=3}p_nnf(n)}{\expval*{n(n-1)}}\geq\sqrt{1-2\frac{\sum_{n=3}^\infty p_nn}{\sum_{n=3}^\infty p_nn^2}} \nonumber\\
    &\quad\times \frac{\sum_{n=3}^\infty p_nn}{\langle n(n-1)\rangle}\left(\frac{\sum_{n=3}^\infty p_nn^2}{\sum_{n=3}^\infty p_nn}-1\right) \nonumber\\
    &\geq \sqrt{1-2\frac{\sum_{n=3}^\infty p_n}{\sum_{n=3}^\infty p_nn}} \frac{\sum_{n=3}^\infty p_nn^2 - \sum_{n=3}^\infty p_nn}{\expval*{n(n-1)}}\nonumber\\
    &=\sqrt{1-2\frac{1-p_0-p_1-p_2}{\expval*{n}-p_1-2p_2}}\left(1-\frac{2p_2}{\expval*{n(n-1)}}\right)
\end{align}
which gives \autoref{eq:general_uncert_limit}. Here we used that $f(x)=\sqrt{1-\frac{2}{x}}$ is monotonically increasing and the Jensen's inequality again such that $\frac{\sum_{n=3}^\infty p_nn^2}{\sum_{n=3}^\infty p_nn} \geq \frac{\sum_{n=3}^\infty p_nn}{\sum_{n=3}^\infty p_n}$. This can be further simplified by noting that $-(1-p_0)\geq-1$ and that $\expval*{n}-p_1\geq2(1-p_0-p_1)$, such that for $\expval*{n}\geq2$ the following inequalities hold
\begin{equation}
    \frac{\expval*{n}-p_1-2p_2}{1-p_0-p_1-p_2} \geq \frac{\expval*{n}-p_1}{1-p_0-p_1} \geq \frac{\expval*{n}-p_1}{1-p_1} \geq \expval*{n}.
\end{equation}
Using again that $f(x)$ is monotonically increasing, we are left with
\begin{equation}
    \Delta\tau\Delta\Omega\geq \sqrt{1-\frac{2}{\expval*{n}}}\left(1-\frac{2p_2}{\expval*{n(n-1)}}\right),
\end{equation}
which is valid for any distribution that fulfills $\expval*{n}\geq2$. This expression can be finally approximated to \autoref{eq:global_bound} by either neglecting the two-photon state probability or, alternatively, assuming that the field is intense enough such that the two-photon probability is much smaller than the average number of photons pairs $p_2\ll\frac{\expval*{n(n-1)}}{2}$.

\end{document}